\documentclass[12pt,twocolumn,authoryear,epsfig]{elsart}
\usepackage{amssymb}
\usepackage{amsmath}
\usepackage{graphicx}
\usepackage{multirow}
\journal{Nuclear Instruments and Methods in Physics Research A}
\begin{document}
\begin{frontmatter}
\title{Measurement of $\beta$-decay end point energy with Planar HPGe detector}
\author[1]{T. Bhattacharjee\corauthref{cor1}},
\corauth[cor1]{Corresponding author}
\ead{btumpa@vecc.gov.in}
\author[1]{Deepak Pandit},
\author[2]{S. K. Das},
\author[1]{A. Chowdhury},
\author[1]{P. Das},
\author[2]{D. Banerjee},
\author[1]{A. Saha},
\author[1]{S. Mukhopadhyay},
\author[1]{S. Pal},
\author[1]{S. R. Banerjee}
\address[1]{Physics Group, Variable Energy Cyclotron Centre, Kolkata, India, PIN - 700 064}
\address[2]{RCD-BARC, Variable Energy Cyclotron Centre, Kolkata, India, PIN - 700 064}

\begin{abstract}

 The end point energies of nuclear $\beta$ decays have been measured with a segmented planar Ge LEPS detector using both singles and coincidence techniques. The $\beta - \gamma$ coincidence has been performed with a segmented planar Ge LEPS and a single 10$\%$ HPGe detector. The $\gamma$ ray and $\beta$ particle responses of the Segmented planer Ge LEPS detector were studied using monte carlo simulation code GEANT3. The experimentally obtained $\beta$ spectrum was in reasonably good agreement with the simulation results. The experimental end point energies are determined with substantial accuracy for some of the known $\beta$ decays in $^{106}$Rh, $^{210}$Bi and $^{90}$Y. The end point energies corresponding to three weak branches in $^{106}$Rh $\rightarrow$ $^{106}$Pd decay has been measured for the first time.\\

\end{abstract}

%

\end{frontmatter}

\section{Introduction}
\label{intro}

The detection of $\beta$ particles has been explored for many years using a variety of detector systems having a wide range of efficiency and resolution, viz., magnetic spectrometer~\cite{ms}, plastic scintillators~\cite{plastic}, silicon~\cite{si} and germanium solid state detectors~\cite{olof,vom}. The magnetic spectrometer provides very high energy resolution but is limited in its efficiency and it is not considered suitable for the measurement involving short lived $\beta$ decaying state. This also requires an elaborate experimental setup as well as very high beam intensities in order to study the $\beta - \gamma$ coincidence connecting the weakly populated states. The plastic scintillators are used for the detection of $\beta$ particles in order to provide a high efficiency detection and fast timing requirements but the poor energy resolution of plastic detectors is overcome by the use of solid state detectors. The silicon detectors have been widely used for this purpose but has a limitation in the detection efficiency specially for very high energy $\beta$ particles, which becomes more abundant as one moves away from the line of stability. The planar Hyper Pure Germanium (HPGe) detectors, having window materials with low Z elements, have already been demonstrated to be quite efficient in this purpose~\cite{rainer,kojima}. Moreover these detectors provide a very good energy resolution as well as can be calibrated up to very high energy using appropriate $\gamma$ ray sources. However, the high atomic number of Ge gives rise to high backscattering and bremsstrahlung production probability making a distortion at the low energy end of the Fermi Kurie plot. In spite of that the high purity germanium detectors are being used in many important fields of high precision nuclear $\beta$ decay~\cite{beta-spiral}.
Experiments with planar HPGe detectors have demonstrated that the end point energies can be determined within a reasonable error by fitting the data points with E $>$ 0.8E$_{\beta}$, E$_{\beta}$ being the maximum energy of the $\beta$ particle~\cite{greenwood}. However, $\beta$ decays far off the $\beta$ stability line become weaker and hence the reasonable fitting of Fermi Kurie plot requires data points ranging a wider energy region.  The fitting of experimental Fermi Kurie plot including the lower energy data points has been demonstrated to incorporate an upward curvature of the fitted line, attributed to the imperfections of the used sources~\cite{greenwood}. Consistent efforts have been given towards the simulations for the $\beta$ response of Ge detectors and understanding the contribution from different sources as a function of energy. The same could be explained well as a combination of several functions corresponding to (i) full energy absorption peak of electron/positron, (ii) the escape of bremsstrahlung photons, (iii) side and back scattering from the detector wall and (iv) the addition of annihilation photon for positrons~\cite{kojima,noma}. Very recently, the GEANT4 code has also been proved to be equally competent to simulate the $\beta$ spectrum generated with a HPGe detector~\cite{new-beta}.

The accurate measurement of the end point energies generated its importance in various fields of nuclear physics. This is used in obtaining the information on Q${_\beta}$  values which in turn determines the atomic mass, one of the most fundamental quantities in nuclear physics. Precise atomic masses, specifically, of the unstable nuclei away from the line of stability, can generate significant inputs to the theoretical model calculations as this manifests all interactions contributing to the nuclear binding~\cite{ref1,ref2}. In recent times, however, different experimental setups have also been available for very precise measurements of atomic masses both for the ground and isomeric states of a nucleus~\cite{ref3,Rb84}.
The experimental determination of $\beta$ decay end point energies play the significant role in the unambiguous establishment of the  level structure of the daughter nuclei. Many of the nuclei around the s-process and rp process path have very long lived $\beta$ decaying isomeric states~\cite{pm148,pm152}. The identification of these isomers is not possible in prompt $\gamma-\gamma$ coincidence study when the half lives range from seconds to hours. Although the decay study of the daughter $\gamma$ rays provides possibilities on their identification via the measurement of decay half lives, the absolute knowledge on the excitation energy and spin parity of such isomers is only possible by the coincidence measurement of the $\beta$ decay of the isomeric state and the corresponding $\gamma$ decay in the daughter nucleus. The $\beta-\gamma$ coincidence measurement with an appropriate $\beta$ detector becomes unique for determining the end point energies related to even very weakly populated levels of the daughter nuclei. This kind of measurements have been employed in some nuclei using plastic scintillators or Si(Li) detectors~\cite{i132}. Studies are known that have employed the $\beta - \gamma$ coincidence technique involving germanium detectors to measure the end point energies~\cite{greenwood}. The application of this experimental technique has been proved to be very important in the study of isomers in neutron rich nuclei~\cite{fogelberg}. Planar HPGe detectors with low Z entrance window are the only choice to measure the weak $\beta$ branches and high energy $\beta$ decays and thus important to explore with the decays having complicated level spectra of the daughter nuclei.

In the present work, the $\beta$ decay end point energies have been determined by using $\beta-\gamma$ coincidence technique for several known and unknown $\beta$ decays. The $\beta$ decay of neutron rich $^{106}$Rh with a complicated $\gamma$ decay scheme of the daughter nucleus $^{106}$Pd has been considered for the measurement. End point energies have been extracted for very weak $\beta$ decay branches by the clean selection through $\gamma$ gating. The experimentally obtained $\beta$ spectrum has been reproduced with substantial accuracy when compared with the results from a Monte Carlo simulation performed with GEANT3 code. The $\chi ^2$ minimisation technique has been employed for known $\beta$ decays, in order to determine end point energy. The said analysis also reflects the confidence limit in the reproduction of the experimental curve by using the GEANT3 simulation. The experimental Fermi Kurie plots for specific $\beta$ branches were constructed in order to determine the end point energies with this conventional technique.
The end point energies of known $\beta$ branches were reproduced with reasonable or better accuracy. The end point energies for few branches of $^{106}$Rh $\beta ^-$ decay have been determined for the first time.

\section{Experimental Setup}
\label{expt}

The experimental setup consisted of a 11mm thick Ge Planar segmented LEPS (Low Energy Photon Spectrometer) detector and a 10$\%$ efficient coaxial HPGe detector. The LEPS detector has a 300 $\mu$m thick Be window which completely stops the $\beta$ particles with energy up to 223 keV and the 10$\%$ Ge detector has response to only very high energy $\beta$ particles for its thick entrance window. The front face of the 10$\%$ detector was covered with 11 mm thick aluminium plate in order to ensure that no $\beta$ particle enters this detector. The open sources of high specific activity with insignificant solid content were prepared and was dried on an electro-polished stainless steel surface of 0.5 mm thickness. The detectors were kept at a distance of 2.9 cm from each other and the source was kept at a distance of 0.9 cm from the LEPS detector. This was done in order to compensate for the different efficiencies of the two detectors. A coincidence measurement was carried out with this setup in order to extract several $\beta$ decay spectra corresponding to the decay of $^{106}$Rh $\rightarrow$ $^{106}$Pd. A standard $^{106}$Ru source was used for this purpose. Singles measurement was carried out only with the LEPS detector for the sources like $^{90}$Sr and $^{210}$Pb, as there are no $\gamma$ transitions with reasonable intensity which are in coincidences with the corresponding $\beta$ decay. The above two sources correspond to the decay paths of $^{90}$Sr $\rightarrow$ $^{90}$Y $\rightarrow$ $^{90}$Zr and $^{210}$Pb $\rightarrow$ $^{210}$Bi $\rightarrow$ $^{210}$Po respectively. Both the singles and coincidence timing electronics were consisted of NIM standard Timing Filter Amplifier (ORTEC 863), Constant Fraction Discriminator (ORTEC 584) and gate and delay generators(ORTEC 8020). For the coincidence setup, prompt coincidence was made within a time window of 50 ns between the CFD `OR' output of the four segments of the LEPS detector and the CFD output from the 10$\%$ HPGe detector by using the four input logic unit (ORTEC CO4020). A Master gate was generated by stretching this coincidence gate in order to incorporate the five energy outputs corresponding to four segments of LEPS and the 10$\%$ HPGe detector. The energy outputs were taken from spectroscopy amplifiers (Canberra 2024) with a shaping time of 4 $\mu$sec for the LEPS detector and 8 $\mu$sec for the other HPGe detector. The data were digitized and acquired with a multiparameter data acquisition system LAMPS~\cite{lamps} and two 13 bit ADCs (ORTEC AD413).

\section{Analysis}
\label{da}

\begin{figure*}
\begin{center}
\hspace{-0.5cm}
\includegraphics[scale=0.65,angle=270]{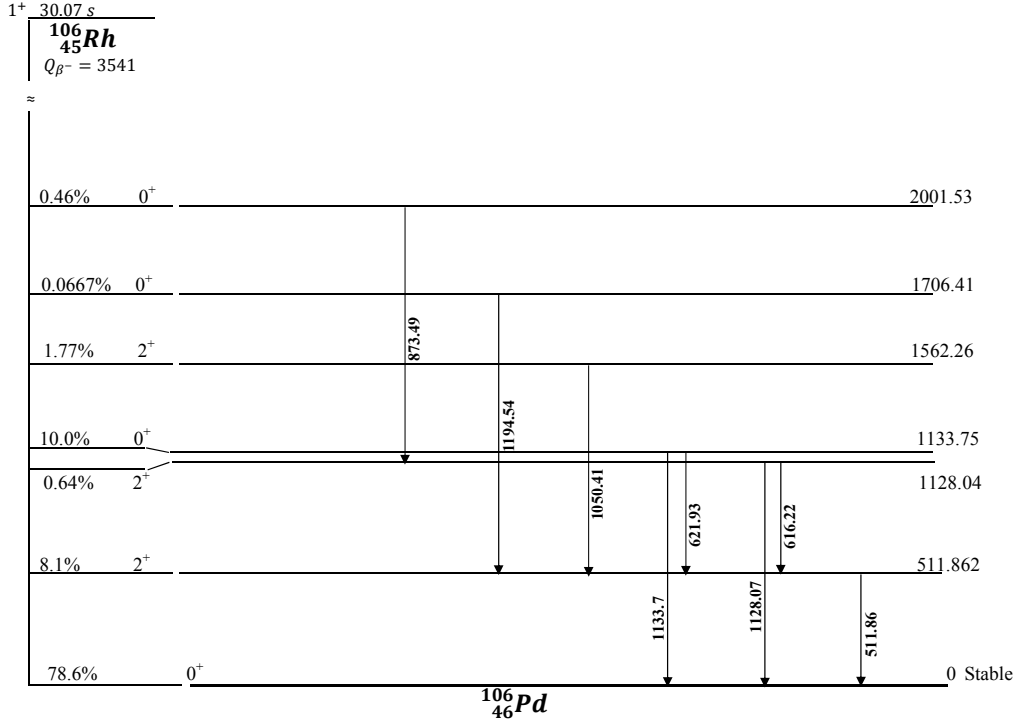}
\caption{Partial decay scheme of $^{106}$Rh $\rightarrow$ $^{106}$Pd , relevant to the present work and taken from ref.~\cite{nndc-rh}, is shown.}
\label{levelscheme}
\end{center}
\end{figure*}

\begin{figure}
\begin{center}
\hspace{-0.5cm}
\includegraphics[scale=0.3,angle=270]{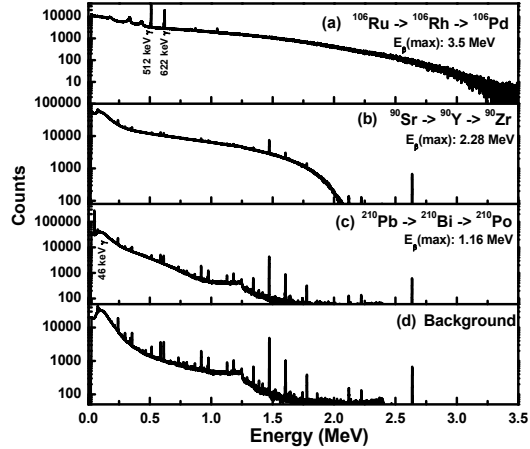}
\caption{Energy spectra obtained with one segment of the segmented planar Ge LEPS detector are shown for different sources.}
\label{spec}
\end{center}
\end{figure}

The decay scheme of the sources $^{90}$Sr and $^{210}$Pb are simpler ones with one high energy and few low energy $\beta$ decay branches~\cite{nndc-sr,nndc-pb}. The decay scheme of $^{106}$Ru followed by the decay of $^{106}$Rh, is a complicated one with a very weak decay branching of 0.64$\%$, 0.07$^\%$ and 0.46$\%$ respectively for the 1128.04, 1706.4 and 2001.5 keV excited levels~\cite{nndc-rh}. The partial decay scheme of $^{106}$Rh $\rightarrow$ $^{106}$Pd, relevant to this work has been shown in Figure~\ref{levelscheme}.
The spectra obtained for these sources with one segment of the LEPS detector have been shown in Figure~\ref{spec}, when no selection was put in any of the detectors. The maximum energy points in each of the spectra corresponds to the Q value of the $\beta$ decay of the parent ground state. For the top spectrum(a), the background $\gamma$ transitions are not visible as the measurement was done in coincidence mode. The next two spectra (b and c) have a good number of $\gamma$ transitions coming from the background activities and this is clearly seen in the background spectrum (d), taken in singles mode. The $\beta$ spectra have been obtained from these raw data as described in the subsection~\ref{spec-beta}. The obtained spectra have been reproduced by using the monte carlo simulation code GEANT3~\cite{geant3} as has been discussed in subsection~\ref{geant3}. In subsection~\ref{endpoint}, the methods have been described for obtaining the end point energies from the corresponding $\beta$ spectrum.

\subsection{Experimental $\beta$ spectra}
\label{spec-beta}

\begin{figure}
\begin{center}
\hspace{-0.4cm}
\includegraphics[scale=0.35,angle=270]{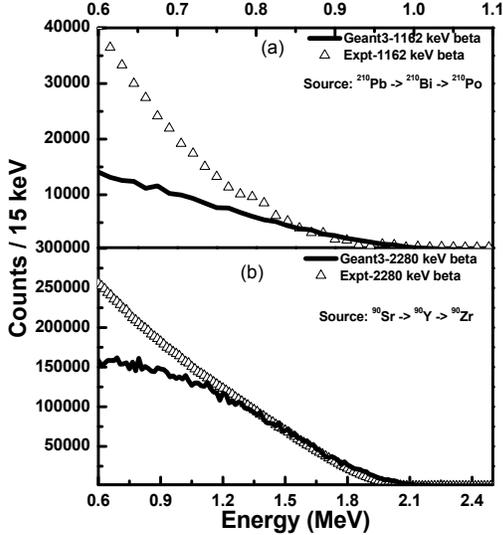}
\caption{$\beta$ spectra obtained for the known decays of (a)$^{210}$Pb $\rightarrow$ $^{210}$Bi $\rightarrow$ $^{210}$Po and (b)$^{90}$Sr $\rightarrow$ $^{90}$Y $\rightarrow$ $^{90}$Zr are shown with $\triangle$. The solid line corresponds to the results from GEANT3 simulation.}
\label{sr-pb-beta}
\end{center}
\end{figure}

\begin{figure}
\begin{center}
\hspace{-1cm}
\includegraphics[scale=0.35,angle=270]{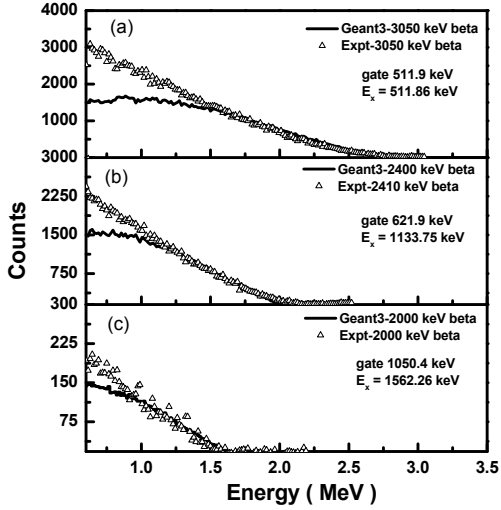}
\caption{$\beta$ spectra obtained for the known decays of $^{106}$Rh $\rightarrow$ $^{106}$Pd are shown with $\triangle$.(a)Decay to 511.86 keV, 2$^+$ level,(b)Decay to 1133.75 keV, 0$^+$ level and (c)Decay to 1562.26 keV, 2$^+$ level of $^{106}$Pd. The solid lines show the corresponding results obtained from GEANT3 simulation.}
\label{ru-known}
\end{center}
\end{figure}

\begin{figure}
\begin{center}
\hspace{-0.4cm}
\includegraphics[scale=0.4,angle=270]{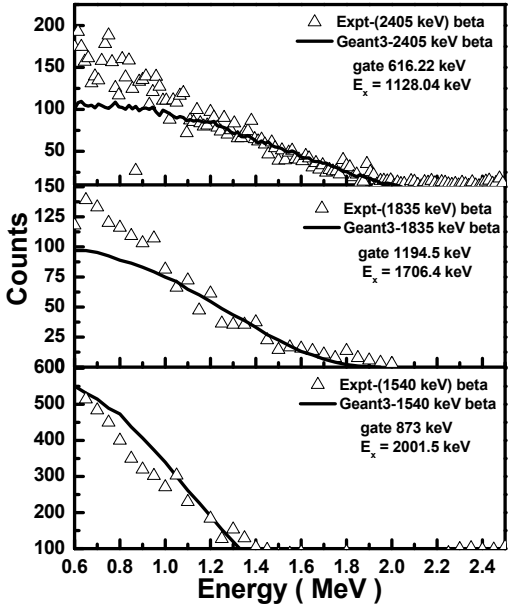}
\caption{$\beta$ spectra obtained for the unknown decays of $^{106}$Rh $\rightarrow$ $^{106}$Pd are shown with $\triangle$. Spectra correspond to the decay to (a)1128.04 keV, 2$^+$ level, (b)1706.41 keV, 0$^+$ level and (c)2001.53 keV, 0$^+$ level of $^{106}$Pd. 15 keV/channel compression for (a), 50 keV/channel compression for (b) and added the spectra from all the segments of the LEPS detector with 50 keV/channel compression for (c) has been used while generating the $\beta$ spectra. The solid lines correspond to the results from GEANT3 simulation.}
\label{ru-new}
\end{center}
\end{figure}

 The $\beta$ spectra of the sources $^{90}$Sr and $^{210}$Pb were extracted by subtracting the background spectrum (Figure~\ref{spec}d) from the corresponding singles spectra(Figure~\ref{spec}b and~\ref{spec}c) after normalizing at one of the strong background $\gamma$ transitions. The $\beta$ spectra thus obtained for the decay of $^{90}$Sr and $^{210}$Pb have been shown in Figure~\ref{sr-pb-beta}. In the decay of $^{210}$Pb, the $\beta$ decay energies are known to be 17.0 keV (84$\%$), 63.5 keV (16$\%$) and 1162.1 keV (100$\%$) whereas in case of the decay of $^{90}$Sr, 546.0 keV (100$\%$), 2280.1 keV (99.98$\%$) and 93.9 keV (1.4$\%$) are the known $\beta$ decay branches. The low energy $\beta$ particles below 223 keV will not be detected by the LEPS detector as will be completely attenuated in the Be window of the detector. Thus, only the high energy decay branches have been indicated in the figure.
 The LEPS detector used in the setup has response for both $\beta$ and $\gamma$ radiations. Hence, the coincidence setup in this work is equipped to detect the coincidences among the $\gamma$ radiations decaying the excited states of daughter nucleus as well as the coincidences among these $\gamma$ radiation and the $\beta$ particles decaying from the ground state of the parent nucleus populating the corresponding excited states of the daughter nucleus. It is thus expected that a coincidence spectrum in the LEPS detector made by gating a $\gamma$ ray, decaying from the level of interest in daughter nucleus, in 10$\%$ HPGe detector, will have contribution from (i.) the respective $\beta$ decay branch populating the level of interest, $\beta_1$ (ii.) the $\gamma$ rays (photopeak and Compton) in coincidence which are decaying from the states below the level of interest, $\gamma_{1i}$s (iii.) the $\gamma$ rays (photopeak and Compton) in coincidence, de-exciting to the level of interest from the higher lying states, $\gamma_{2j}$s  and (iv.) the $\beta$ branch decaying to these higher lying states, $\beta_2$. Hence, it is important to properly delineate and subtract all other contributions and to extract the required $\beta$ spectrum. For this purpose the coincidence measurements were taken in two configurations, viz., (a) the open source facing the LEPS detector and putting no absorber other than the factory made Be window in front of LEPS and (b) the open source facing the 10$\%$ Ge detector and keeping a 8 mm thick Ta block in front of the LEPS detector. $\gamma$ ray gate was put in the 10$\%$ detector and the corresponding spectra were projected in four segments of the LEPS detector. The spectrum from configuration (a) consists of the response both from $\beta$ and $\gamma$ radiations whereas the spectrum from configuration (b) consists of response only from $\gamma$ in the LEPS detector. The spectra were normalized at the most intense photopeak in coincidence and the second spectrum was subtracted from the first one in order to arrive at the required $\beta$ spectrum. The contribution from the unwanted low energy $\beta$ branches, feeding the higher lying levels, were subtracted by appropriately subtracting the two different $\gamma$ gates. While generating the $\beta$ spectra the original spectra were compressed in order to increase the statistical accuracy. 15 keV/ channel binning was employed for strong decays and 50 keV/channel binning were done for weak decay branches. With this method the $\beta$ spectra have been obtained for several known and unknown branches of $^{106}$Rh $\rightarrow$ $^{106}$Pd decay. The spectra obtained for three known branches, viz. to the 511.86 keV (2$^+$) level, to the 1133.75 keV (0$^+$) level and to the 1562.26 keV (2$^+$) level of $^{106}$Pd, have been shown in Figure~\ref{ru-known}.
The $\beta$ spectra corresponding to three weak and unknown branches of $^{106}$Rh decay, viz. to 1128.04 keV (2$^+$) state, to 1706.41 keV (0$^+$) state and 2001.53 keV (0$^+$)state, have been shown in Figure~\ref{ru-new}. The known values of the $\beta$ decays have been indicated in Figure~\ref{ru-known} along with the energy of the excited state to which the decay ends. The decay energy values indicated in Figure~\ref{ru-new} corresponds to the difference in Q value and the energy of the excited level to which the decay ends. The gating transition has also been indicated in all the figures. In order to extract the $\beta$ spectra for known decays and the unknown decay to the 1128.04 keV (2$^+$) level of $^{106}$Pd, a 15 keV/channel compression has been used and the spectra have been generated from the single crystal of the LEPS detector.  In case of the decay to the 1706.41 keV, 0$^+$ level of $^{106}$Pd a 50 keV/channel compression has been used to increase the statistics. The $\beta$ decay spectrum corresponding to the decay to 2001.53 keV, 0$^+$ level has been obtained by adding the spectra from all the segments of the LEPS detector and a compression of 50 keV/channel.
In the Figures~\ref{sr-pb-beta},\ref{ru-known} and~\ref{ru-new}, the triangles show the extracted data points and the solid line show the results obtained with the monte carlo simulation GEANT3. The results have been explained in detail in the following subsection~\ref{geant3}.

\subsection{Monte Carlo simulation with Geant3}
\label{geant3}

\begin{figure}
\begin{center}
\includegraphics[scale=0.3,angle=270]{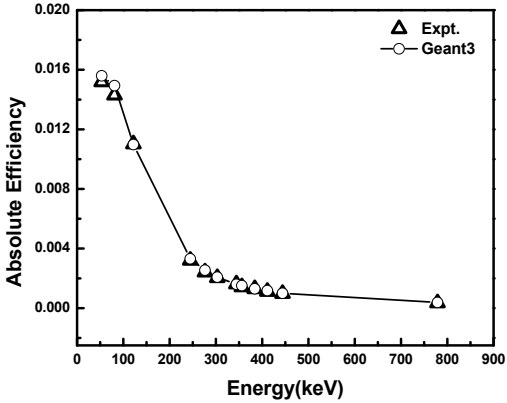}
\caption{Efficiency of $\gamma$ rays obtained with one segment of the segmented planar Ge LEPS detector.}
\label{eff-gamma}
\end{center}
\end{figure}

A Monte carlo simulation has been performed for all these $\beta$ energy spectra in the LEPS detector by Geant3 simulation package~\cite{geant3}. The relevant physics processes ionization, multiple scattering and bremsstrahlung for electrons, while photoelectric effect, Compton scattering and pair production for gamma rays were included in the simulation. The simulations were performed with exact geometrical condition taking into account the germanium crystal, the beryllium window and the aluminium casing. In order to confirm the proper consideration of the detector geometry in the simulation the absolute efficiency of the LEPS detector have been measured for $\gamma$ detection. The measurement has been performed with standard sources like $^{152}$Eu and $^{133}$Ba having known disintegration per second(dps). The experimental results have been shown in Figure~\ref{eff-gamma} and it matches very well with the simulated efficiency, shown in the same figure. The simulation for different $\beta$ decay branches have then been performed and plotted with the corresponding experimental results in the Figures~\ref{sr-pb-beta}, ~\ref{ru-known} and~\ref{ru-new}. The $\beta$ decay energies have been considered as the difference in the Q value and the energy of the excited level of the daughter nucleus to which the decay takes place. It is observed that the simulation could reproduce almost all the $\beta$ spectra except their low energy part. At low energy the simulation underestimates the experimental data points. Other than the known reasons of source imperfections this is assumed to be due to the background obtained from the $\beta$-Compton and Compton-Compton coincidence events that could not be subtracted while producing the $\beta$ spectra by coincidence method. In the spectra obtained in singles measurement, the mismatch is more compared to the coincidence measurement because of the huge Compton background underlying the spectra that could not be properly subtracted. A possible inaccuracy in simulation at low energies could also arise from the backscattering coefficients as observed in ref~\cite{new-beta}. The mismatch in the higher energy part beyond the end point energy is conjectured to be due to the underlying background which can be estimated as a second order polynomial in energy.

\subsection{Determination of end point energy}
\label{endpoint}
\begin{table*}[ht!]
\begin{center}
\caption{The details of studied $\beta$ decays with the obtained endpoint energies in the present work.}
\begin{tiny}
\label{table1}
\begin{tabular}{|c|cc|c|ccccc|}
\hline
\hline
Decaying&\multicolumn{2}{c|}{Final Level}&Branching&\multicolumn{5}{c|}{$\beta$ Energy}\\
Nucleus&Energy&J$^{\pi}$&&(Q$_{\beta}$ - E$_x$)&Lit. Val.&\multicolumn{2}{c}{F-K Anal.}&GEANT3\\
&&&&&&(uncorr)&(corr)&and\\
&&&&&&&&$\chi ^2$ Anal.\\
(Q$_{\beta}$)&(E$_x$)(keV)&&($\%$)&(keV)&(keV)&(keV)&(keV)&(keV)\\
\hline
$^{106}$Rh $\rightarrow$ $^{106}$Pd&511.9&2$^+$&8.1&3029.1&3050$\pm$20&2836$\pm$40&2926$\pm$40&3045$\pm$15\\
(3541.0 keV)&&&&&\cite{nndc-rh}&&&\\
&1133.8&0$^+$&10.0&2407.3&2410$\pm$30&2232$\pm$34&2319$\pm$34&\\
&&&&&\cite{nndc-rh}&&&\\
&1562.3&2$^+$&1.77&1978.7&2000$\pm$100&1842$\pm$65&1928$\pm$65&1987$\pm$22.5\\
&&&&&\cite{nndc-rh}&&&\\
&1128.0&2$^+$&0.64&2412.9&-&2150$\pm$103&2246$\pm$103&\\
&1706.4&0$^+$&0.067&1834.6&-&1898$\pm$208&1984$\pm$208&\\
&2001.5&0$^+$&0.46&1539.5&-&1764$\pm$317&1862$\pm$317&\\
$^{210}$Bi $\rightarrow$ $^{210}$Po&0.0&0$^+$&100.0&1162.1&1161.5$\pm$15&&&\\
(1162.1 keV)&&&&&\cite{nndc-pb}&&&\\
$^{90}$Y $\rightarrow$ $^{90}$Zr&0.0&0$^+$&99.98&2280.1&2275$\pm$3&&&\\
(2280.1 keV)&&&&&\cite{nndc-sr}&&&\\
\hline
\hline
\end{tabular}
\end{tiny}
\end{center}
\end{table*}

\subsubsection{End point energy from the Fermi-Kurie Analysis}
\label{fk}

\begin{figure}
\begin{center}
\includegraphics[scale=0.28,angle=270]{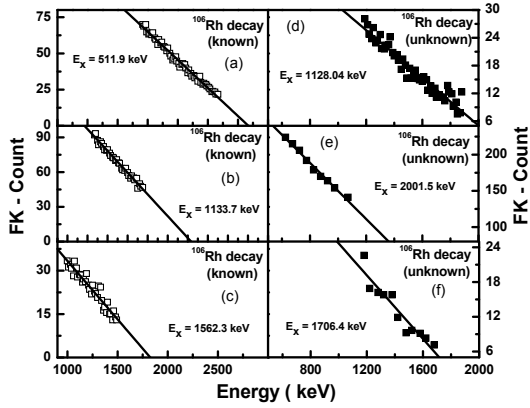}
\caption{Fermi-Kurie plot for different $\beta$ decays of $^{106}$Rh $\rightarrow$ $^{106}$Pd. (a), (b), (c) are known $\beta$-decays and shown with open squares. (c), (d) and (e) are unknown $\beta$-decays and are shown with filled squares. Corresponding excited levels of the daughter nucleus is mentioned in the plots.}
\label{fkplot}
\end{center}
\end{figure}

The Fermi-Kurie plots were generated from the obtained $\beta$ spectrum by using the built in routine FK-Energy in spectrometric handbook in the spectrometer code LISE~\cite{lise,kantele}. The Fermi-Kurie plot corresponding to different $\beta$ branches have been shown in Figure~\ref{fkplot}. The end point energies have been obtained from the corresponding Fermi-Kurie plot by fitting the data points with a linear function $y=a.x+b$. The end point energies has been obtained from the relation $E_p = b/a + d$, where d is the correction factor for the degradation at the window material of the planar detector. The end point energies, both uncorrected and corrected, obtained in the present work have been tabulated in Table~\ref{table1} and compared with the literature value. In the present work, the known $\beta$ decays of $^{106}$Ru, $^{90}$Sr and $^{210}$Pb have been measured with reasonable accuracy and the $\beta$ decay energies for three weak branches in $^{106}$Ru decay have been determined for the first time.

\subsubsection{End point energy from the GEANT3 simulation and Chi-Square Analysis}
\label{cs}

\begin{figure}
\begin{center}
\includegraphics[scale=0.3,angle=270]{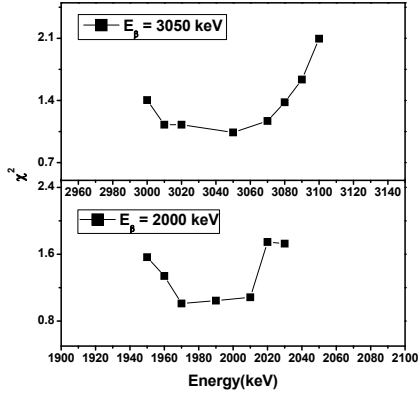}
\caption{\label{chi-square}$\chi ^2$ spectra for two known $\beta$ decays of $^{106}$Rh.}
\end{center}
\end{figure}
In this technique, the GEANT3 simulation has been performed for obtaining the $\beta$ decay spectra by varying the end point energies up to $\pm$100 keV from the known value of a particular beta decay end point energy. Input beta particle spectra with different end point energies around the expected value was projected on the detector and the beta spectra from the detector was generated. A chi square analysis has then been performed with these simulation results and the experimentally obtained data, for the known $\beta$ decays of $^{106}$Rh. The $\chi^2$ has been obtained by standard techniques by comparing the experimental result with the same from different simulations, as given in the following equation.
\begin{equation}
\chi ^2 = \frac{1}{N-1}\sum _{i=1}^{N} \frac {(d_i -t_i)^2}{\sigma}
\end{equation}
where the N is the total number of data points, d$_i$ is the experimental value and t$_i$ are the corresponding values obtained with simulation. $\sigma$ is the standard error which has been calculated as,
\begin{equation}
\sigma = \sqrt{d_i}*1.2
\end{equation}
Since error in the detected number of beta particles cannot be simple statistical error, we have considered other possible error contributions by enhancing the statistical error bar by 20$\%$.
The $\chi^2$ values have been plotted as a function of the considered end point energies and the results have been shown in Figure~\ref{chi-square} for two different known $\beta$ decays. The energy value that corresponds to the munimum $\chi^2$ has been considered as the obtained end point energy. The uncertainty in the obtained end point energy has been considered as the energy variation for which the $\chi^2$ increases by 10$\%$ from the minumum value. The end point energies have been obtained for two known beta decays of $^{106}$Rh and are tabulated in Table~\ref{table1}.

\section{Discussion}
\label{dis}

The measurements on $\beta$ decays have been performed for $^{210}$Bi, $^{90}$Y and several decay branches of $^{106}$Rh using segmented planar Ge LEPS detector with 300 $\mu$m Be window. $\beta$ spectra have been generated by subtracting the $\gamma$ background from singles and coincidence measurements. Experimental data have been reproduced by monte carlo simulation performed with GEANT3 code. The end point energies have been calculated by using the Fermi-Kurie distribution as well as comparison of the experimental data with the simulated ones by calculating the $\chi ^2$ values. End point energies have been obtained for the first time for three weak $\beta$ decay branches in $^{106}$Rh $\rightarrow$ $^{106}$Pd decay.
\section{Acknowledgement}
\label{ack}
 The authors acknowledge the sincere efforts of Mr. R. K. Chatterjee who have assisted meaningfully in the preparation of the sources. The authors have been benefitted by the efforts of the members of physics lab in maintaining the Ge detectors throughout the year.

\end{document}